# Multi-Parametric Statistical Method for Estimation of Accumulated Fatigue by Sensors in Ordinary Gadgets

Nikita Gordienko,
Physical-Mathematical Lyceum 142, Kyiv, Ukraine
assasin.nik@gmail.com

Abstract

The new method is proposed to monitor the level of currently accumulated fatigue and estimate it by the several statistical methods. The experimental software application was developed and used to get data from sensors (accelerometer, GPS, gyroscope, magnetometer, and camera), conducted experiments, collected data, calculated parameters of their distributions (mean, standard deviation, skewness, kurtosis), and analyzed them by statistical methods (moment analysis, cluster analysis, bootstrapping, periodogram and spectrogram analyses). The hypothesis 1 (physical activity can be estimated and classified by moment and cluster analysis) and hypothesis 2 (fatigue can be estimated by moment analysis, bootstrapping analysis, periodogram, and spectrogram) were proposed and proved. Several "fatigue metrics" were proposed: location, size, shape of clouds of points on bootstrapping plot. The most promising fatigue metrics is the distance from the "rest" state point to the "fatigue" state point (sum of 3 squared non-normal distribution of non-correlated acceleration values) on the skewness-kurtosis plot. These hypotheses were verified on several persons of various age, gender, fitness level and improved standard statistical methods in similar researches. The method can be used in practice for ordinary people in everyday situations (to estimate their fatigue, give tips about it and advice on context-related information).

1.Background

The standard cardiology monitoring can show the instant state of cardiovascular system, but unfortunately, cannot estimate the accumulated fatigue and physical exhaustion. From 2009 FIFA persists that professional players should record family history, heart rhythm, sounds, and ECG (electrocardiogram) results [1]. Despite this more than 60 professional football players died while playing a game or training of a suspected heart attack or cardiac arrest during the last decade only! On May 6, 2016, Patrick Ekeng (football team Dinamo Bucureşti), 26, collapsed in a match 7 minutes after came on from the bench and died less than two hours later of a suspected heart attack. Before the game he told his best friend he was not able to play, he said he was very tired [2]. The common reason for such sad statistics is the increased physical load and overtraining that lead to dangerous fatigue and fatal outcome. In addition to sport, fatigue could be crucial in many other areas (car driving, air and naval traffic control, manufacture and service industry), where errors due to fatigue can lead to decrease of working efficiency, manufacturing quality, and, especially, workplace and customer safety. That is why people need some easily accessible ways to estimate their fatigue and physical exhaustion. There are some specialized commercial accelerometers, which are used to record the number of steps, etc [3-4]. However, they are quite primitive in terms of data, and are not able to assess the health state and measure fatigue [5-8].

2.Experimental Procedures and Methods of Statistical Analysis

The main aim of this research is to test theoretical possibility of monitoring the health of individuals based on physical data gathered by sensors in usual smartphones (and analysis of the data) for the assessment of human fatigue and potential prediction of signs of some diseases, especially for the elderly.

To reach this aim the following tasks were performed: explore the possibility of using sensors in conventional gadgets (smartphone, smart watches, smart glasses) to monitor fatigue with their analysis of the physical activity; develop and apply methods for measuring basic physical quantities that characterize human mobility (acceleration, speed, etc.); create and use multi-parametric methods of statistical analysis, which use several parameters of distributions (mean, standard deviation, skewness and kurtosis) of the physical data obtained from sensors of conventional gadgets; offer practical ways for implementation of these methods.

Experimental methodology is based on frequent measurements (every 1-100 milliseconds) acceleration of the movements of certain parts of the body by the axes X, Y, Z via G-sensor (acceleration sensor or accelerometer) in a conventional mobile phone, or in a more convenient, but rare smart watches or smart glasses. The amplitude acceleration values vary very widely for different types of activity and operations. And it is very hard to find some patterns even for very different activities. However, if you plot the distribution of physical quantities, their difference is quite noticeable. For numerical characteristics of this difference the standard parameters of distributions were selected: mean value, standard deviation, skewness, and kurtosis. The statistical analysis of experimental data was performed for distributions of acceleration values over a short period of time (for 1-10 minutes), then the standard parameters of the distributions were calculated and correlations with types of physical activity were determined.

3.Results

3.1.Analysis of physical activity based on 2 and 3 parameters of distribution

Each point on the graphs below (Fig.1-2) corresponds to parameters of some distribution of large number of measurements. Each point contains a label of the body part, where sensors were located. For example, WALK_LEG_L - indicates measurement during walking (WALK), a sensor was located on the left leg (LEG_L). It was noted that the parameters of the distribution are monotonically dependent on the speed (Fig.1-2). This means that the motor activity (of species listed in small text at the characters) can be classified according to the intensity that is divided into groups (colored ovals) with close values of parameters distributions: active behavior (sports, housework, walking - highlighted in blue) moderate behavior (letter, seat - marked in green) and passive behavior (web surfing, reading, sleeping - highlighted in red) (Fig. 13-15). Thus, it is possible to characterize and classify the level of human mobility by means of the cluster analysis.

| 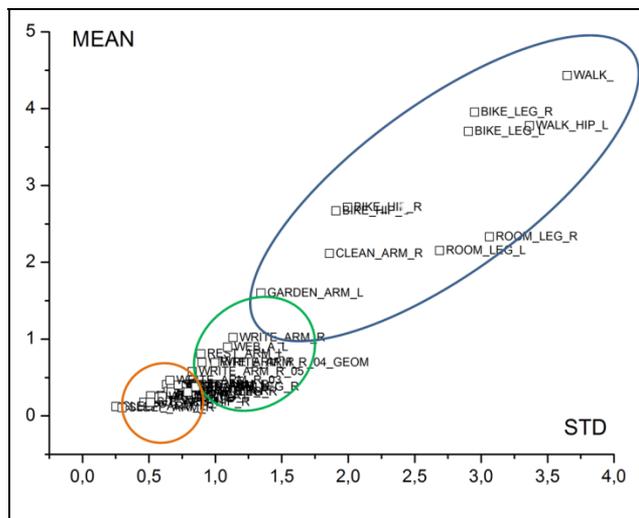 | 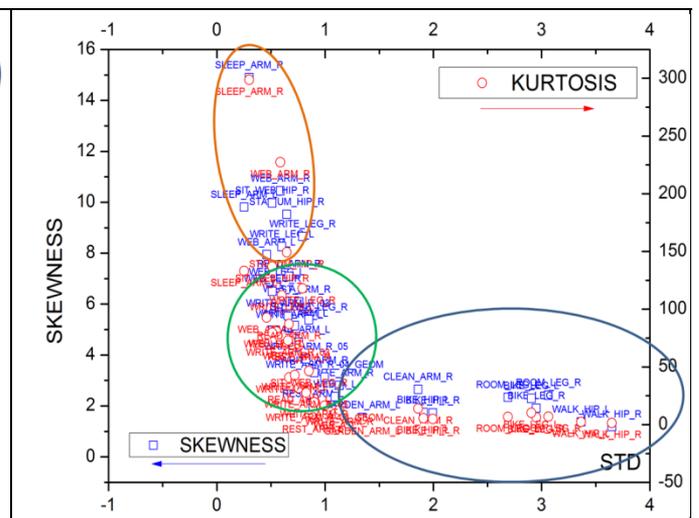 |
|---|---|
| Fig.1.Mean (MEAN) and standard deviation (STD) for various distributions, corresponding to different physical activities and location of sensors. | Fig.2.Skewness, kurtosis, and standard deviation (STD) for various distributions, corresponding to different physical activities and location of sensors. |

That is why the statistical analysis by 3 parameters of distributions can be more reliable and accurate for determination of the level of physical activity. These results are in good correspondence with the previous similar experiments on measurements of physical activities by accelerometers [9].

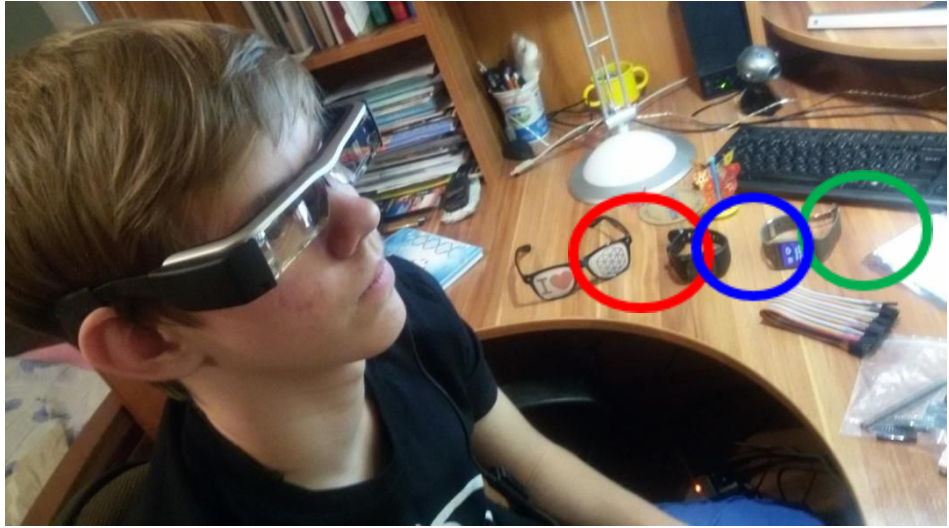

Fig.3.Smart glasses EPSON Moverio BT-200 (on the head) as a collector of data; Conventional glasses with Smart Particle Photon controller with an accelerometer (in the left red circle); Texas Instruments watches with accelerometer (the medium blue circle); smart watches Samsung Gear 2 with accelerometer (the right green circle).

3.2.Analysis of fatigue based on 2 and 3 parameters of distribution

To estimate fatigue, the interrelation of the muscle power to the mass of the body part is very important. The lower muscle power and higher the mass of a body part, the more pronounced effect of fatigue in the shape of tremor. For example, tremor can be measured by subtle shakes of head or fingers of outstretched hands. So the smart glasses and ordinary glasses with Particle Photon controller with accelerometer were used to test the manifestations of fatigue from shaking head (Fig.3).

The same multi-parametric statistical analysis was applied for assessment of human fatigue after physical exercises (squats for a minute) by smart glasses and ordinary glasses (Fig.3). The measurements were made every 1 millisecond (Fig.4) while sitting (letters and ellipse around them with index 1) and standing for physical exercise (with index 2), and while standing (index 4) and seat (index 5) after exercise. The parameters after exercise differ from values before the exercise, especially at the initial moment when people have not had time to recover.

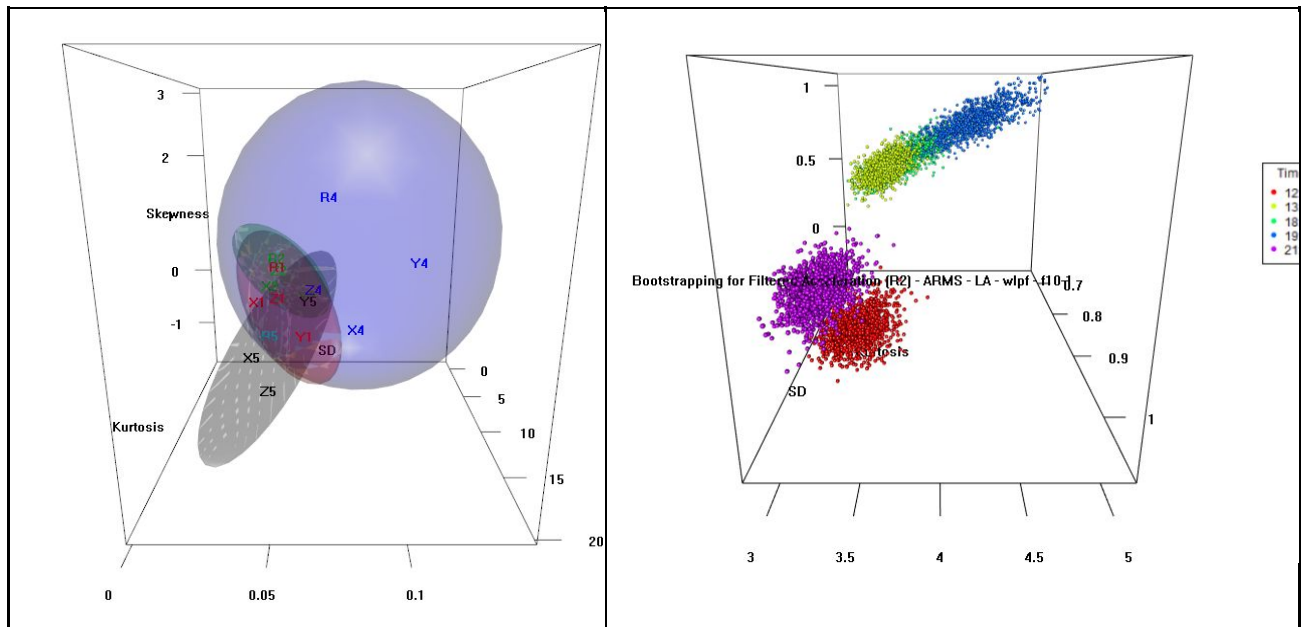

| Рис.4.Groups of parameters of distributions enclosed by ellipses for individual stages of the experiment (the explanations of symbols are given in the text). | Fig.5.Parameters of partial samplings for tremor of outstretched hands with a smartphone in the hands in the rest states (compact clouds) and fatigue states (elongated clouds) |

### 3.3.Analysis of partial sampling (bootstrapping)

To analyze the stability of the obtained distribution parameters the method of partial sampling (or bootstrapping method) from the original distribution of acceleration values. In Fig.5 each point represents one of > 1000 partial samples from the original distribution (hand tremor of the housewife, 45 years) with the correspondent parameters: the standard deviation (axis is directed from the depths of the page), the skewness (vertical axis) and kurtosis (horizontal axis). Different colors mean different time of measurements (which is indicated in the legend). In Fig.5 the qualitative difference is clearly seen: compact equiaxial (nearly round) "clouds" of dots correspond to vigil state (morning and evening) and elongated "cloud" - tired state (after 10 km ski walk). The most promising fatigue metrics is proposed as the distance from the "rest" state point, which is actually position of Chi-square distribution (sum of 3 squared normal distribution of non-correlated acceleration values), to the "fatigue" state point (sum of 3 squared non-normal distribution of non-correlated acceleration values) on the skewness-kurtosis plot.

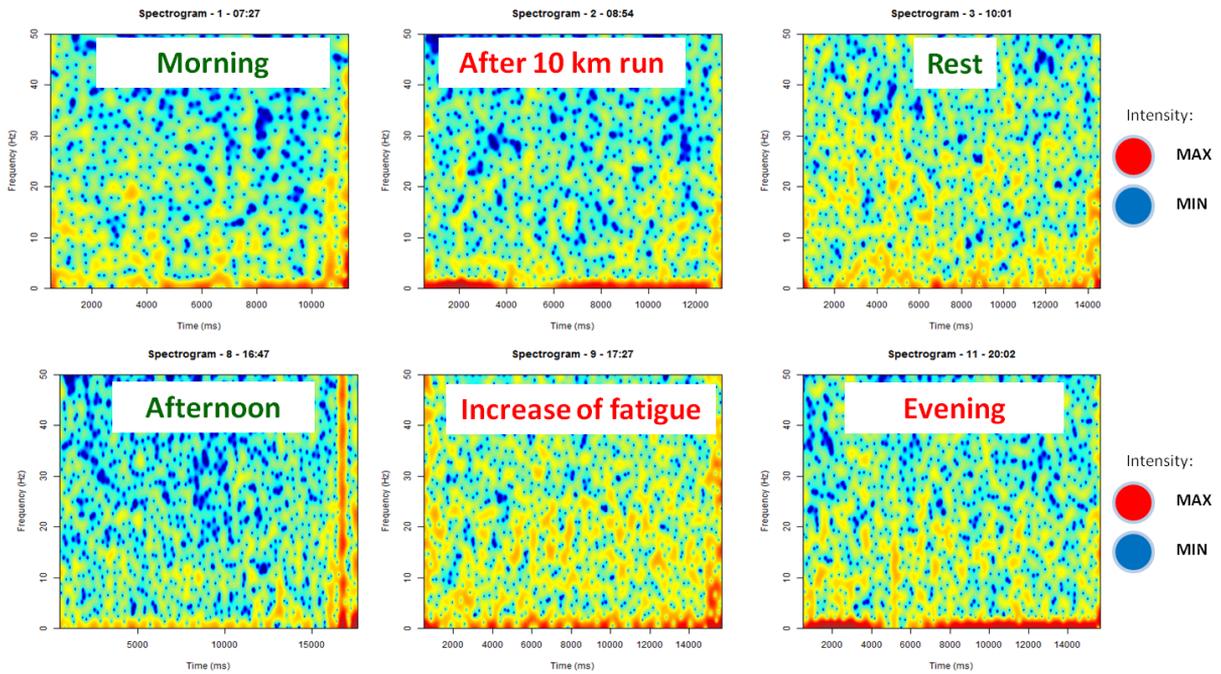

Fig.6.Spectrogram of tremor of outstretched hands with a smartphone.

3.4.Spectral analysis

The spectral analysis (i.e. amplitudes for the range of frequencies for a given measurement interval 12-15 seconds) is shown in Fig.6 for tremor of outstretched hands with a smartphone during the day. The amplitude is denoted by colors from blue (lowest) to red (highest). Slow increase of fatigue manifests itself as an increase of the amplitude (red area at the bottom of the spectrogram) of uncontrolled low-frequency tremor: after the morning run 10 km (time 8:54), fatigue accumulation up to the evening (17:27) and the greater fatigue at late night (20:02). The slow relaxation (i.e. no explicit maximum of uncontrolled low-frequency tremor) is observed during the day except for the evening time. For example, the largest fatigue levels are observed after the morning run of 10 km, before the delayed lunch and in the evening!

4.Conclusions

The experimental results obtained during measurements of body part accelerations by ordinary and new gadgets allow to propose the new method to monitor the level of currently accumulated fatigue and estimate it by the several statistical methods. The experimental software application was developed and used to get data from sensors (accelerometer, GPS, gyroscope, magnetometer, and camera), conducted experiments, collected data, calculated parameters of their distributions (mean, standard deviation, skewness, kurtosis), and analyzed them by statistical methods (moment analysis, cluster analysis, bootstrapping, periodogram and spectrogram analyses). The hypothesis 1 (physical activity can be estimated and classified by moment and cluster analysis) and hypothesis 2

(fatigue can be estimated by moment analysis, bootstrapping analysis, periodogram, and spectrogram) were proposed and proved. Several "fatigue metrics" were proposed: location, size, shape of clouds of points on bootstrapping plot. The most promising fatigue metrics is the distance from the "rest" state point, which is actually position of Chi-square distribution (sum of 3 squared normal distribution of non-correlated acceleration values), to the "fatigue" state point (sum of 3 squared non-normal distribution of non-correlated acceleration values) on the skewness-kurtosis plot. These hypotheses were verified on several persons of various age (16-49), gender (M/F), fitness level (child, housewife, and … marathoner even) and improved standard statistical methods in similar researches. The method can be used in practice for ordinary people in everyday situations (to estimate their fatigue, give tips about it and advice on context-related information).